\RequirePackage{lineno}
\RequirePackage{lineno}
 \documentclass[prd,twocolumn,preprintnumbers,superscriptaddress,nofootinbib]{revtex4}
\usepackage{tabularx} 
\usepackage[title]{appendix}
\usepackage{graphicx}  
\usepackage{dcolumn}   
\usepackage{bm}        
\usepackage{amssymb}   
\usepackage{multirow}
\usepackage{lineno}
\usepackage{wrapfig}
\usepackage[hyperfootnotes=false]{hyperref}
\begin{document}
\newcommand{\beq}{\begin{equation}}
\newcommand{\eeq}{\end{equation}}
\newcommand{\de}{\delta}
\newcommand{\di}{\displaystyle}
\newcommand{\ep}{\epsilon}
\newcommand{\ga}{\gamma}
\newcommand{\Ga}{\Gamma}
\newcommand{\la}{\lambda}
\newcommand{\om}{\omega}
\newcommand{\si}{\sigma}
\newcommand{\ve}{\varepsilon}
\newcommand{\vp}{\varphi}
\newcommand{\dmbarbest} {{(3.36^{+0.46}_{-0.40} {(stat.)}\pm0.06{(syst.)})\times 10^{-3} eV^{2}}}
\newcommand{\dmbest}  {(2.32^{+0.12}_{-0.08})\times10^{-3} eV^{2} }
\newcommand{\pk}{{k \cdot p}}
\newcommand{\pkprime}{{k' \cdot p}}
\newcommand{\kq}{{k \cdot q}}
\newcommand{\pq}{{q\cdot p}}
\newcommand{\W}{{p'}}
\newcommand{\qW}{{q \cdot p'}}
\newcommand{\Wq}{{q \cdot p'}}
\newcommand{\pW}{{p\cdot p'}}
\newcommand{\GeV}{\; {\mathrm{GeV}}}
\newcommand{\cm}{\; {\mathrm{cm}}}
\newcommand{\D}{\displaystyle}

\newcommand{\minerva}{MINER$\nu$A}
\newcommand{\ttbs}{\char'134}
\newcommand{\integ}{\iint}
\newcommand{\FA}{${\cal F}_A$}
\newcommand{\fa}{${\cal F}_A(q^2)$}
\newcommand{\gepQ}{$G_{Ep}(q^2)$}
\newcommand{\nubar}[0]{$\overline{\nu}$}
\newcommand{\gep}{G_{Ep}}
\newcommand{\gmp}{G_{Mp}}
\newcommand{\gmn}{G_{Mn}}
\newcommand{\gmpmu}{G_{Mp}/\mu_{p}}
\newcommand{\gepK}{G^{Kelly}_{Ep}}
\newcommand{\gmpK}{G^{Kelly-upd}_{Mp}}
\newcommand{\gen}{G_{En}}
\newcommand{\gmnmu}{G_{Mn}/\mu_{n}}
\newcommand{\gepnew}{G_{Ep}^{new}}
\newcommand{\gmpnewmu}{G_{Mp}^{new}/\mu_{p}}
\newcommand{\gmpKmu}{G^{Kelly-upd}_{Mp}/\mu_{p}}
\newcommand{\gennew}{G_{En}^{new}}
\newcommand{\gmnnewmu}{G_{Mn}^{new}/\mu_{n}}
\newcommand{\numu}{\nu_{\mu}}
\newcommand{\muminus}{\mu^{-}}
\newcommand{\muplus}{\mu^{+}}
\newcommand{\numubar}{\overline{\nu}_{\mu}}
%
%
\newcommand{\carbon}{\rm ^{12}C}
\newcommand{\oxygen}{\rm ^{16}O}
\newcommand{\deuteron}{\rm ^{2}H}
\newcommand{\hydrogen}{\rm ^{1}H}
\newcommand{\Hefour}{\rm ^{4}He}
\newcommand{\lead}{\rm^{208}Pb}
\newcommand{\Hethree}{\rm ^{3}He}
\newcommand{\neon}{\rm ^{20}Ne}
\newcommand{\aluminum}{\rm^ {27}Al}
\newcommand{\calcium}{\rm^ {40}Ca}
\newcommand{\argon}{\rm ^{40}Ar}
\newcommand{\iron}{\rm ^{56}Fe}
\newcommand{\genie}{$\textsc{genie}$}
\newcommand{\qv}{$\bf |\vec q|$}
\newcommand{\rlqe}{${\cal R}_L^{QE}(\bf q, \nu)$ }
\newcommand{\rtqe}{${\cal R}_T^{QE}(\bf q, \nu)$ }
\newcommand{\rltot}{${\cal R}_L(\bf q, \nu)$ }
\newcommand{\rttot}{${\cal R}_T(\bf q, \nu)$ }
 \newcommand{\rlsimple}{${\cal R}_L$ }
\newcommand{\rtsimple}{${\cal R}_T$ }
\newcommand{\Rochester}{Department of Physics and Astronomy, University of Rochester, Rochester, NY  14627, USA}
\newcommand{\JLAB}{Thomas Jefferson National Accelerator Facility, Newport News, VA 23606, USA}
\title{Parameterizations of  Electron Scattering Form Factors for  Elastic Scattering and Electro-Excitation of Nuclear States for  $\rm ^{27}Al$ and $\rm ^{40}Ca$ } \author{Arie~Bodek}
\affiliation{\Rochester}
\email{bodek@pas.rochester.edu}
   \author{M.~E.~Christy}
\affiliation{\JLAB}
\email{christy@jlab.org}
\author{Zihao Lin}
\affiliation{\Rochester}
\email{zlin22@ur.rochester.edu}
\author{Giulia-Maria  Bulugean}
\affiliation{\Rochester}
\email{gbulugea@ur.rochester.edu}
\author{Amii Daniela Matamoros Delgado }
\affiliation{\Rochester}
\email{amatamor@u.rochester.edu}
\date{\today}
\begin{abstract}
We report on empirical parameterizations of  longitudinal (${\cal R}_L$) and transverse (${\cal R}_T$) nuclear  electromagnetic form factors for elastic scattering and the excitations
of nuclear states  in ${\rm ^{27}Al}$ and ${\rm ^{40}Ca}$. The parameterizations are needed for the calculations of radiative corrections in measurements of electron  scattering cross sections on ${\rm ^{27}Al}$ and ${\rm ^{40}Ca}$ in the  quasi-elastic, resonance and inelastic continuum regions,  provide the contribution of nuclear excitations in investigations of the Coulomb Sum Rule, and test
 theoretical model predictions for excitation of nuclear states in electron and neutrino interactions on nuclear targets at low energies.
\end{abstract}

\maketitle
%
\section{Introduction}
\label{intro}
 Precision measurements of the scattering of electrons from nuclear targets require the application of radiative corrections. Precise calculations of radiative corrections require modeling of electron scattering cross sections as a function of energy, angle and energy transfer. These include elastic scattering of electrons from the nucleus,  nuclear excitations, quasi-elastic scattering including meson exchange currents, resonance production and inelastic scattering processes. However, many of the calculations of radiative corrections do not include the cross sections for excitation of nuclear states.
 \\
 
 For example, the cross section for electro-excitation of   $\rm^{27}Al$ nuclear states are needed in the measurement of parity-violating asymmetry in electron scattering measurements by the Qweak collaboration at Jefferson Lab.  This measurement is used in the  determination of the neutron distribution radius in $\rm ^{27}Al$~\cite{Qweak:2021ijt}, and for the $\rm ^{27}Al$ contribution to the measurement of the proton weak charge\cite{Magee:2016rva}. Similarly, the calculations of  radiative correction in  precision measurements of the longitudinal and transverse electromagnetic structure functions of $\rm ^{12}C$, $\rm^{27}Al$ and $\rm^{56}Fe$ by the JUPITER collaboration\cite{JUPITER:2025uny} also require cross sections for the electro-excitation of nuclear states for these nuclei. Consequently,  we embarked on a program to parameterize the form factors for elastic scattering and nuclear excitations for various nuclei. 
\\
 
\begin{figure}[ht]
\begin{center}
\includegraphics[width=3.45in,height=2.6in]{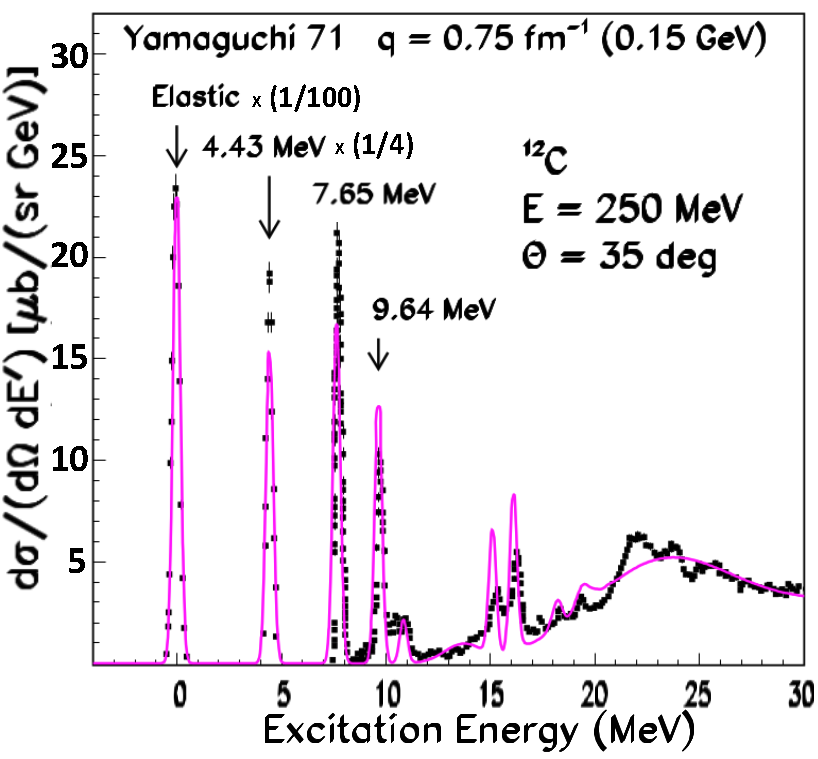}
\caption{ An example from our previous analysis of $\rm ^{12}C$ nuclear excitations: Radiatively corrected cross section from Yamaguchi:71\cite{Yamaguchi:1971ua}(measured with high resolution of 0.25\%) for the scattering of 250 MeV electrons from  ${\rm ^{12}C}$ at 35$^0$.  Here, the cross section for the elastic peak has been divided by 100, and the cross section for the 4.43 MeV state by 4.5}
\label{C12_states_1}
\end{center}
\end{figure}

 In a previous communication \cite{Bodek:2023dsr}, we reported on empirical parameterizations of the form factors for elastic scattering and the excitation of nuclear states for $\rm ^{12}C$. These are used in the recent
 precision extractions\cite{JUPITER:2025uny,Alsalmi:2019sie} of the longitudinal and transverse electromagnetic structure functions of $\rm ^{12}C$.  A comparison of an electron scattering cross section measurement\cite{Yamaguchi:1971ua} on $\rm ^{12}C$ to our empirical parameterization is shown in Fig. \ref{C12_states_1}.

%

 These $\rm ^{12}C$ parameterization were recently used in the calculation of radiative corrections of precision measurements of the transverse and longitudinal structure functions of  $\rm ^{12}C$\cite{JUPITER:2025uny}. They have  also been  used in evaluation of the contribution of nuclear excitations to the Coulomb sum rule \cite{Bodek:2022gli} for $\rm ^{12}C$.
 
 In this communication we report on the parameterizations of the form factors for elastic scattering and the excitation of nuclear states with excitation energy less than 10 MeV for $\aluminum$ and $\calcium$.
We note that a significant number of form factor measurements were not available in tabular form and required digitization from the figures in the publications.  A recent analysis of only the elastic form
factors for $\aluminum$ and $\calcium$ can be found in \cite{Noel:2024led}.
%
%
\begin{figure*}[ht]
\begin{center}
\includegraphics[width=3.3in,height=3.0in]
{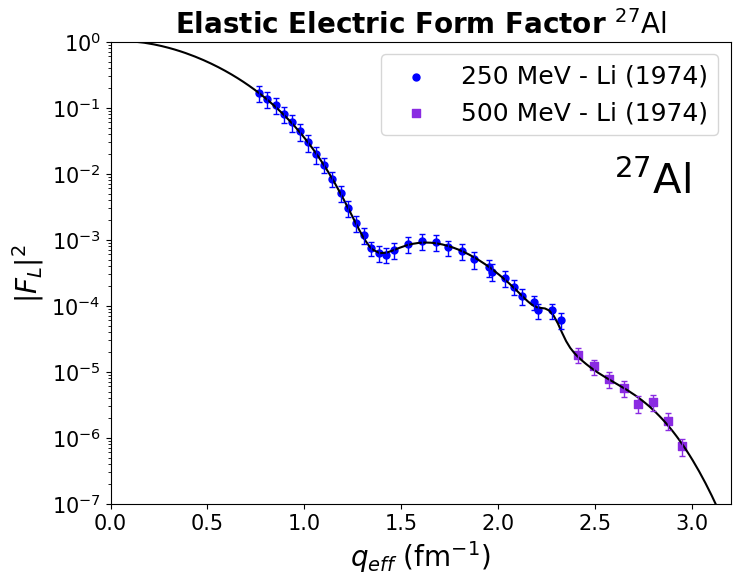}
\includegraphics[width=3.4in,height=3.05in]
{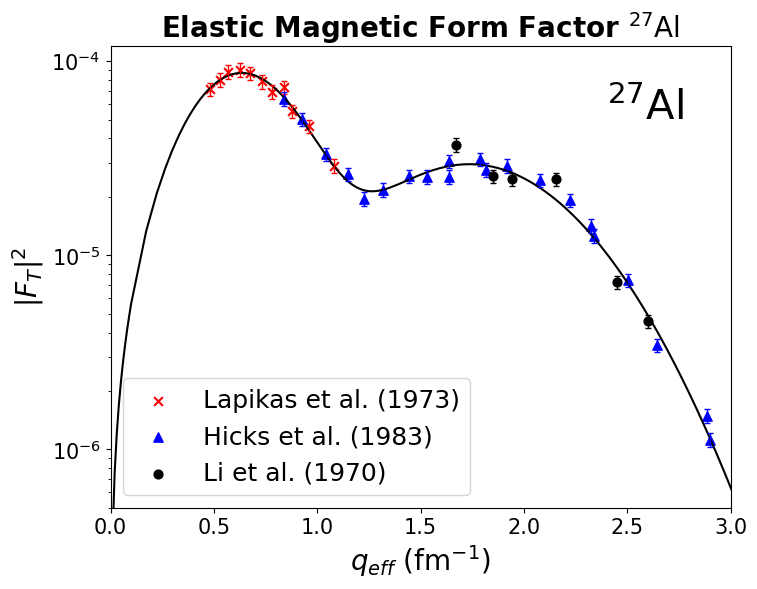}
\caption{Nclear elastic electric and magnetic form factors squared for $\aluminum$. Error bars that are not visible are within the points. The data are taken from
\cite{Li:1974vj}, \cite{Donnelly:1984rg} and \cite{Lapikas:1973njq}.}
\label{AL27_elastic}
\end{center}
\vspace{-20 pt}
\end{figure*} 

\subsection{Electron scattering from nuclear targets}
\label{concepts}
For incident  electrons of  energy $E_0$, scattered electron energy $E^{'}$, and  scattering angle  $\theta$, the electron scattering cross section for elastic scattering and excitation of nuclear states can be written as \cite{Li:1974vj}:
\begin{eqnarray}
\frac{d\sigma}{d\Omega}&=&Z^2 \sigma_M \frac{1}{\eta}[ \frac{Q^4}{\bf{ q^4}} |F_L(\bf{q^2)}|^2 \nonumber \\
&+& (\frac{Q^2}{2 \bf{q^2}} +\tan^2\frac{\theta}{2})  |F_T(\bf{q^2)}|^2  )],
\label{sigma_mott}
\end{eqnarray}
where 
\begin{eqnarray}
\eta&=&1 + 2\frac{E_0}{M_A}\sin^2 \theta/2~(for~elastic~ scattering), \nonumber\\
\eta&=&1~(for~excitation~of~nuclear~states), \nonumber\\
Q^2 &=&  4E_0E^{'}{\sin}^2 \frac{\theta}{2},
\end{eqnarray}
and the square of the magnitude of  3-momentum transfer vector $\vec {\bf q}$ is
\begin{equation}
{\bf q}^2 = Q^2 +\nu^2.
\end{equation}
The Mott cross section  $\sigma_M$ is given by 
\begin{eqnarray}
\sigma_M &= & \frac{\alpha^2 \cos^2(\theta/2) }{[2 E_0  \sin^2(\theta/2)]^2}  = 
 \frac{4\alpha^2E^{\prime 2}}{Q^4} \cos^2(\theta/2).\\
\end{eqnarray}
Here $M_A$ is the mass of the nuclear target,  $M$ is the mass of the proton and  the energy transfer to the target  $\nu = E_0-E^{\prime}$.  The mass  of  the $\aluminum$ nucleus is   $M_{A=27}$ =   26.982 u = 25.133 GeV. The mass  of  the $\calcium$ nucleus is   $M_{A=40}$ =   39.963 u = 37.225 GeV (1 u = 931.502 MeV). In these expressions we have neglected  the electron mass which is negligible for the kinematics studied. 
For scattering from a nuclear target the excitation energy $E_x$ is given by $E_x=\nu-\nu_{elastic}$  where
\begin{equation}
\nu_{elastic}= E_0-\frac{E_0}{1+2E_0{\sin}^2 \frac{\theta}{2}/M_A}.
\end{equation} 
Or equivalently 
\begin{equation}
E_x=\nu-\frac{Q^2_{elastic}}{2M_A},
\end{equation}
where  $Q^2_{elastic}$ is $Q^2$ for elastic scattering from a carbon nucleus for incident energy $E_0$ and
scattering angle $\theta$.  

%
%
The squares of the form factors for elastic scattering  and nuclear excitations are parameterized as a function of the effective momentum transfer
\cite{Oguro:1984zz,Wang:2005gi} $\bf q_{eff}$ where   
\begin{eqnarray}
q \equiv  |\bf {q_{eff}}| &=& |\bf {q} (1+\frac{3Z\alpha}{2E_0\ R})|.
\label{q_eff}
\end{eqnarray}

Here R=3.94 fm  For  ${\rm ^{27}_{13}Al}$\cite{Ryan:1983zz}, and R=4.5 fm for ${\rm ^{40}_{20}Ca}$\cite{Oguro:1984zz}. Using equation \ref{q_eff}, we determine the multiplicative correction to $\bf{q}$ for  0.19 GeV incident energy on ${\rm ^{27}_{13}Al}$ is   1.0375, and the correction for 0.25 GeV incident energy on ${\rm ^{27}_{13}Al}$ is 1.0285.   
%
\section {Parameterizations of Nuclear Elastic and Nuclear Excitations form factors}
\subsection {$\aluminum$  elastic form factors}
Since the  $\aluminum$ 
 nucleus has spin 5/2, it  has both magnetic (transverse)  and  electric (longitudinal, charge)  elastic form factors. 

The square of the  elastic nuclear electric (charge) form factor for ${\rm ^{27}Al}$ is parameterized by:
\begin{eqnarray}
 F_c^2(\bf {q_{eff}}^2)^L&=&H^2(\bf {q_{eff}}^2)+\sum_{j=1}^{j=3}  N_j e^{-[(\bf {q_{eff}}^2-C_j)/\sigma_j]^2} \nonumber \\
 &+ & \frac{\bf {q_{eff}}}{\bf {q_{eff}} +A_L} N_4e^{-[(\bf {q_{eff}}^2-C_4)/\sigma_4]^2} ,\nonumber \\
 H(\bf {q_{eff}})& =&[1-\frac{\alpha \bf {q_{eff}}^2 a_0^2}{2(2+3\alpha)}] exp[\frac{-\bf {q_{eff}}^2 a_0^2}{4}], 
  \end{eqnarray}
  where  $\bf {q_{eff}}$ is the effective momentum transfer in units of fm$^{-1}$, $A_L$= 0.038, and  
 $H({\bf {q_{eff}}^2)}$ is the harmonic well shape (with $\alpha$= 1.504, and $a_0$=2.116 for $\aluminum$).
 \vspace{10 pt}

  $N_1= 5.973\times 10^{-4}$, $C_1=2.179$, $\sigma_1=1.576$.
  
  $N_2=3.525\times 10^{-5}$, $C_2=5.111$, $\sigma_2=0.277$.
  
  $N_3=5.529\times 10^{-6}$, $C_3=6.284$, $\sigma_3=1.721$.
  
  $N_4=7.771\times 10^{-2}$, $C_4=0.004$, $\sigma_4=0.708$.
\vspace{10 pt}

    The square of the elastic nuclear  magnetic  form factor is parameterized using the following expression,
    \begin{equation}
F^2_M (\bf {q_{eff}})^T=  \frac{\bf {q_{eff}}}{\bf {q_{eff}} +A_M} \times \sum_{j=1}^{j=3}  N_j e^{-[(\bf {q_{eff}} -C_j)/\sigma_j]^2},
\end{equation}
 where  $A_M=0.381$, and   
 
 $N_1=1.394\times 10^{-4}$, $C_1=0.585$, $\sigma_1=0.379$,
 
  $N_2=3.587\times 10^{-5}$, $C_2=1.716$,   $\sigma_2=0.649$, 
  
  $N_3=8.803 \times 10^{-8}$, $C_3=1.716$,   $\sigma_3=0.650$.
\\

The $A_M$ and $A_L$ terms serve to adjust parameterizations to improve the  fit values at lower {\bf{q}}. 

In the parameterizations of the elastic nuclear electric form factors squared, the Gaussians are parametrized versus the square of the effective 3-momentum transfer $\bf {q_{eff}^2}$, while for the nuclear elastic magnetic  form factor as well as for  all electro-excitation of nuclear states the Gaussians are parametrized versus the magnitude of the effective 3-momentum transfer $\bf {q_{eff}}$.

 A comparison of our fit for the squares of the  elastic  nuclear   form factors of $\aluminum$ to experimental data is shown in Fig.  \ref{AL27_elastic}.
  The left panel shows the square of the elastic electric form factor versus $\bf {q_{eff}}$ in fm$^{-1}$.
 The right panel shows the square of the  elastic  magnetic form factor versus $\bf {q_{eff}}$ in fm$^{-1}$.
  %
\begin{figure}[h!]
  \centering
  \includegraphics[width=3.5in,height=3.6in]{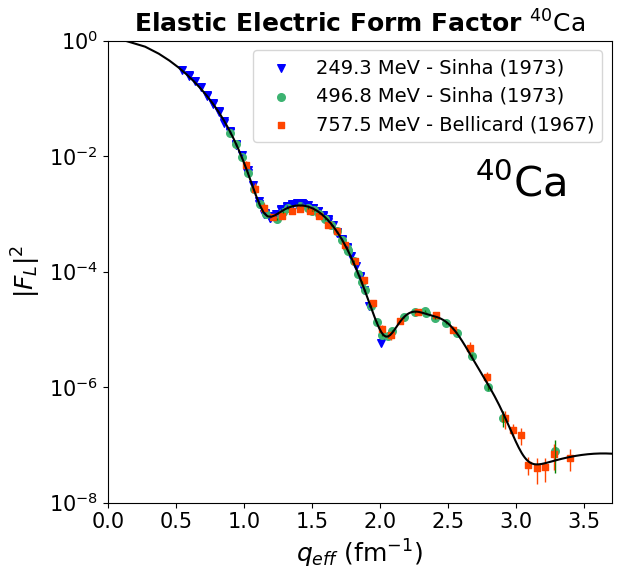}
  \includegraphics[width=3.45in,height=2.5in]{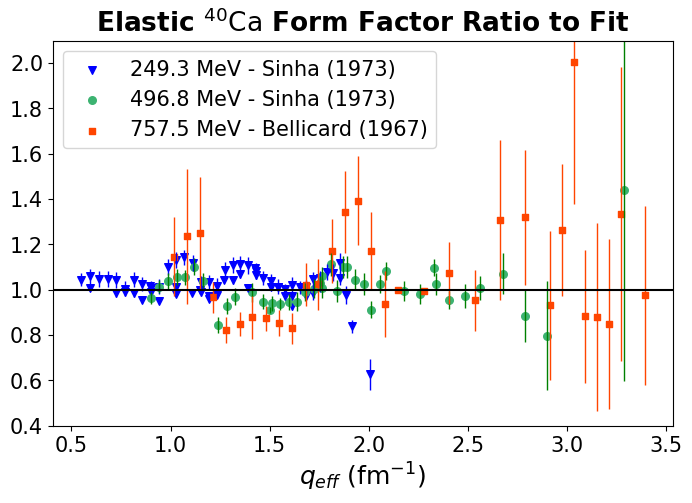}
\caption{{\bf{Top panel:}} The square of the elastic  nuclear electric form factor for $\calcium$ versus $\bf {q_{eff}}$. Measurements taken from \cite{Sinha:1973zz} and \cite{PhysRevLett.19.527}. {\bf {Bottom panel:}} Ratio of $F_c^2(\bf {q_{eff}}^2)^L$ to the fit.}
\label{ca40_elastic}
\vspace{-20 pt}
\end{figure}

\begin{figure*}[ht]
\begin{center}
\includegraphics[width=4.3in,height=1.6in]{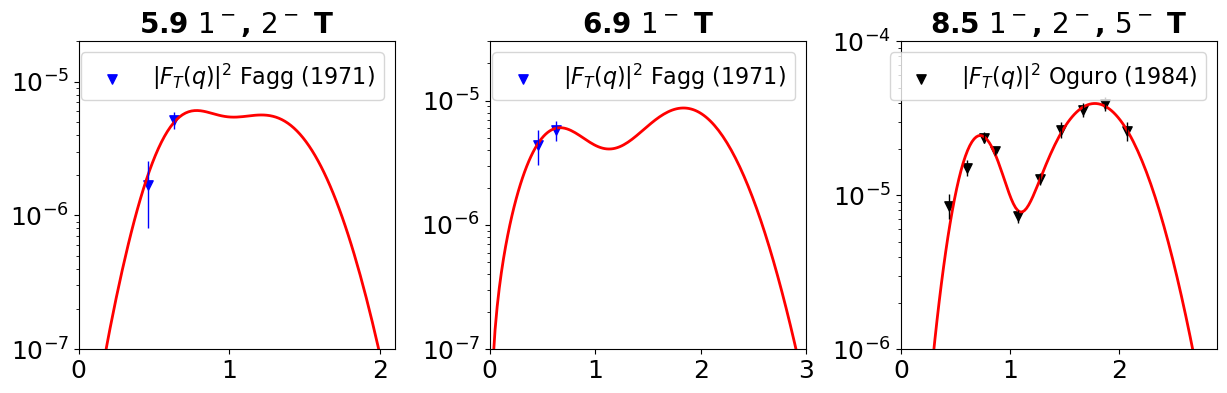}
\includegraphics[width=2.7in,height=1.7in] {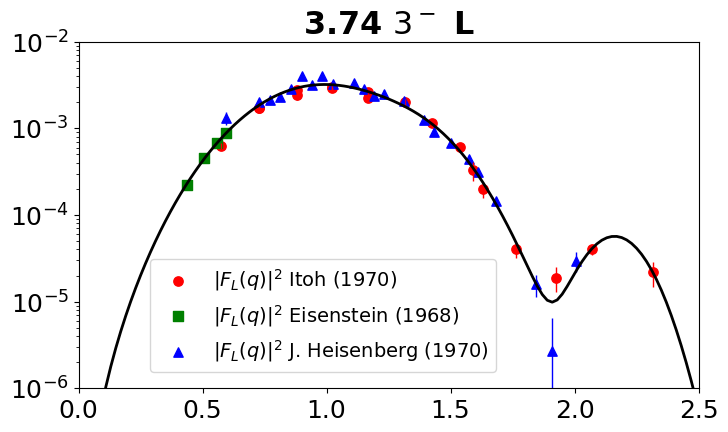}
\includegraphics[width=4.3in,height=1.8in] {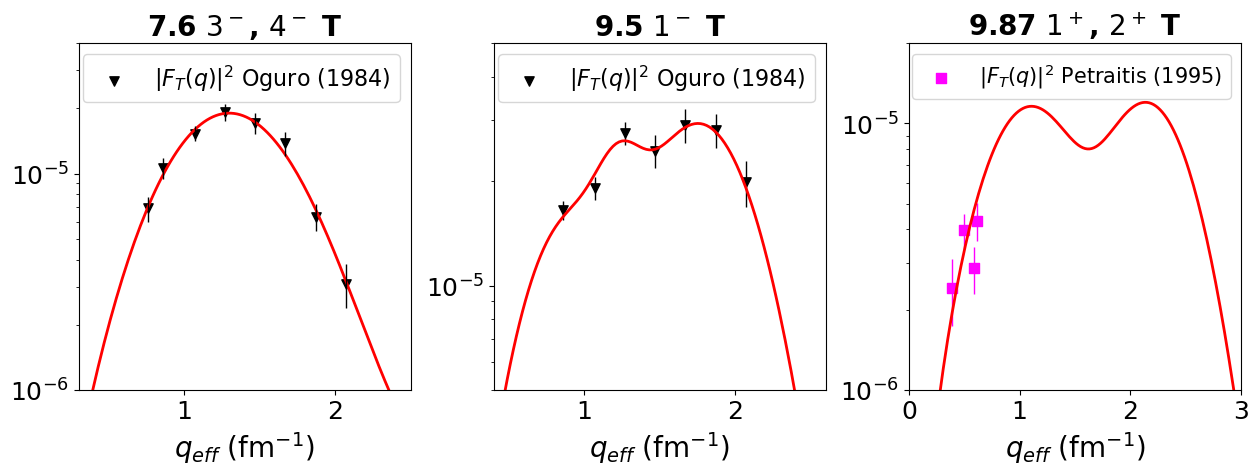}
\includegraphics[width=2.5in,height=1.6in] {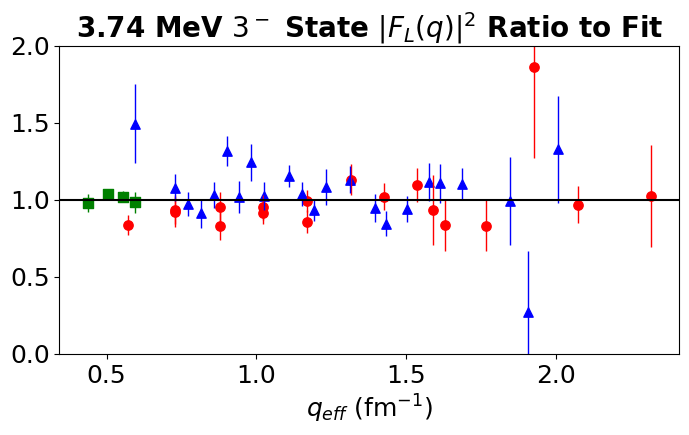}
\caption{{\bf{Top and Bottom Left}}: Comparison of measurements of $\calcium$ transverse nuclear excitation form factors (squared)  with the parameterizations detailed in  Table \ref{excited_states_ca40T}. Data taken from \cite{Petraitis:1995zz} (Petraitis:1995), \cite{Oguro:1984zz} (Oguro:1984), and \cite{Fagg:1971zz} (Fagg:1971). {\bf{Top Right}}: Longitudinal Form factor for the 3.74 MeV odd parity nuclear excitation state. Data taken from \cite{Itoh:1970jb}, \cite{Heisenberg:1971rw}, and \cite{PhysRev.188.1815}. Parametrization (black solid line) using  parameters in Table \ref{excited_states_ca40L}. {\bf{Bottom Right}}: Ratio for the 3.74 MeV $F_L^2$ to the fit.}
\label{40Ca_sample_states}
\end{center}
\end{figure*}
%
%
\subsection {$\calcium$  elastic nuclear electric  form factor}
We fit the measured $\calcium$ elastic nuclear electric  form factor with the following form (similar to the one used to parameterize the $\aluminum$ elastic nuclear electric form factor): 
\begin{eqnarray}
 F_c^2(\bf {q_{eff}}^2)^L&=&H^2( \bf {q_{eff}}^2)+\sum_{j=1}^{j=6}  N_j e^{-[(\bf {q_{eff}}^2-C_j)/\sigma_j]^2},\nonumber \\
 H(\bf {q_{eff}}^2)& =&[1-\frac{\alpha  \bf {q_{eff}}^2  a_0^2}{2(2+3\alpha)}] exp[\frac{-\bf {q_{eff}}^2a_0^2}{4}],
  \end{eqnarray}
 where $H(\bf{q_{qeff}}^2)$ is the harmonic well shape (with $\alpha$= 1.256, and $a_0$=2.595, for $\calcium$). 
\\

The  $\calcium$ elastic electric form factor values are extracted from the $\calcium$  elastic scattering cross-section measurements\cite{Sinha:1973zz,PhysRevLett.19.527} 
using equation \ref{sigma_mott} (with  $|F_T(\bf{q^2)}|^2$ = 0), and posterior equations defined in section A. 
Comparison of the parametrization of the square of the  nuclear elastic electric form factor for $\calcium$ to experimental data is shown on the top panel on Fig \ref{ca40_elastic}. The ratio of experimental data to the fit is shown in the bottom panel. 
\\

The Gaussian parameters are listed below:

  $N_1= 1.138\times 10^{-3}$, $C_1=1.996$, $\sigma_1=0.825$.
  
  $N_2=1.383\times 10^{-5}$, $C_2=5.758$, $\sigma_2=0.908$.
  
  $N_3=1.248\times 10^{-5}$, $C_3=4.911$, $\sigma_3=0.568$.
  
  $N_4=2.337\times 10^{-6}$, $C_4=6.593$, $\sigma_4=1.308$.
  
  $N_5=7.079\times 10^{-8}$, $C_5=1.325$, $\sigma_5=4.616$.

  $N_6=8.932\times 10^{-2}$, $C_6=0.112$, $\sigma_6=0.522$.
  %
  
\begin{figure*}[ht]
\begin{center}
\includegraphics[width=7.2in,height=2.2in] {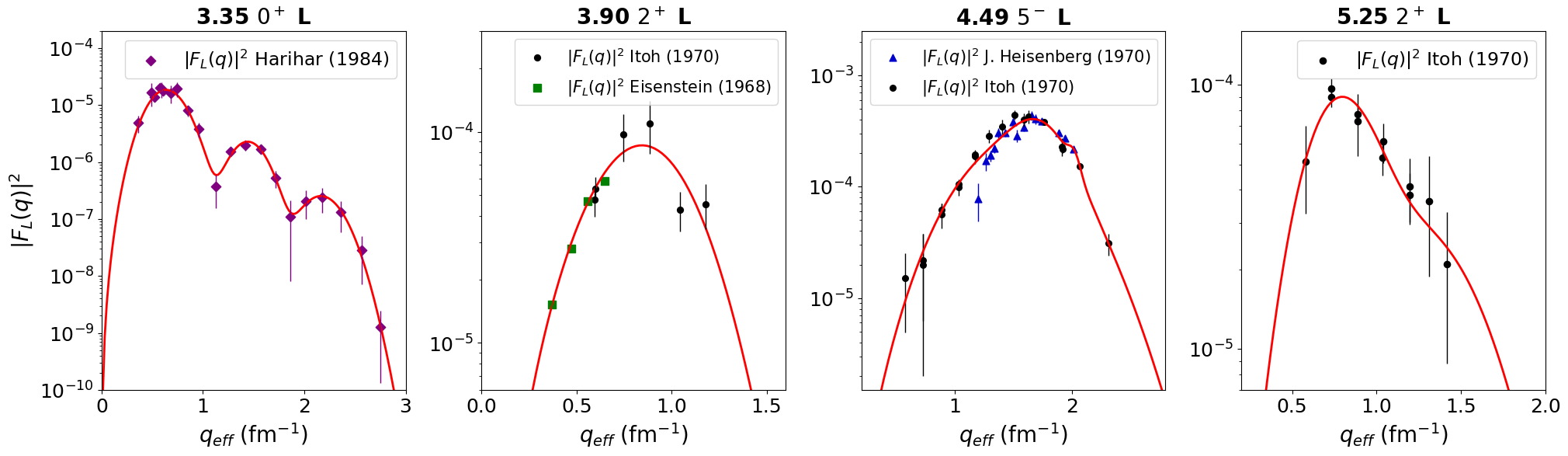}
\includegraphics[width=7.2in,height=2.0in] {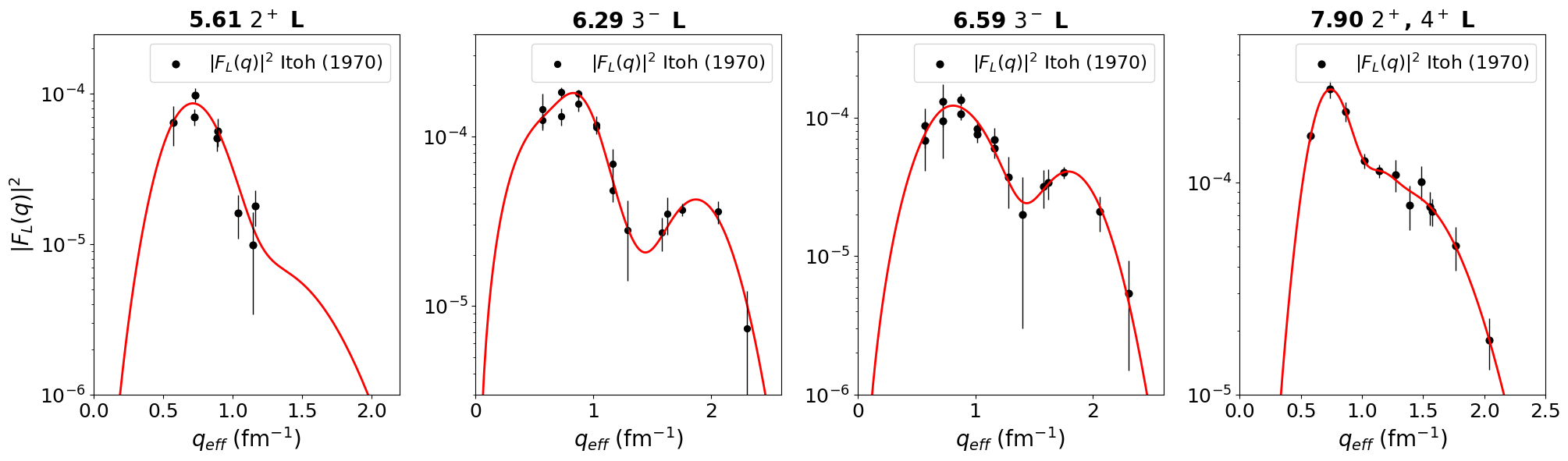}
\includegraphics[width=6.2in,height=2.0in]{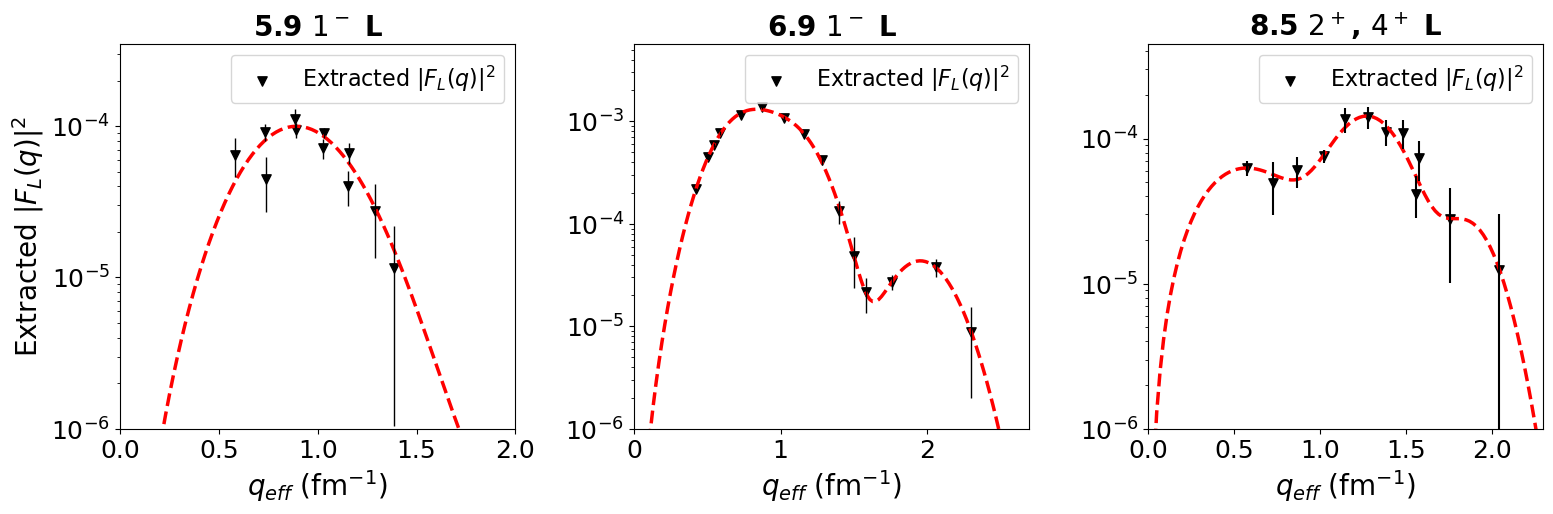}
\caption{{\bf{Top:}} Longitudinal nuclear excitation  form factors for $\calcium$. Parametrizations (solid red line) given by the parameters in Table \ref{excited_states_ca40L}. Measurements from  \cite{Heisenberg:1971rw} (blue triangle), \cite{Itoh:1970jb} (black dots), and \cite{PhysRevLett.53.152} (purple diamond). {\bf{Bottom:}} Extracted $|F_L (q)|^2$ $\rm^{40}Ca$ nuclear excitation form factors from experimental total $|F(q)|^2$ measurements from \cite{Itoh:1970jb}, in combination with $|F_T (q)|^2$ measurements from  \cite{Fagg:1971zz}. Parameterizations shown in dotted red line.}
\label{40Ca_statesL}
\end{center}
\vspace{-20 pt}
\end{figure*}
%
\setlength\extrarowheight{5pt}
\begin{table*}[ht]
\begin{center}
\begin{tabular}{|p{22mm}||p{16mm}|p{11mm}|p{13mm}||p{16mm}|p{11mm}|p{13mm}||p{16mm}|p{11mm}|p{14mm}|}
        \hline
\multicolumn{10}{|c|}{\textbf{Parameters of our fits of $\calcium$ transverse nuclear excitation form factors.}} \\ \hline
State	&	$N_1$	&	$C_1$	&	$\sigma_1$		&	$N_2$ &	$C_2$	&	$\sigma_2$	&	$N_3$ &	$C_3$	&	$\sigma_3$	 \\  \hline \hline
5.90 1-,2- T	& 	4.800E-06 &	0.660	& 0.250 &	3.800E-06 &	0.850 &	0.450 &	2.800E-06 &	1.600 & 0.380 \\ \hline
6.90 1- T	&  	6.600E-06 &	0.500 &	0.350 & 3.800E-06 & 0.700 & 0.700 &	4.500E-06 & 1.800 &	0.500 \\ \hline
7.6  3-,4-T	& 1.488E-05 &	1.169 &	0.567 & 1.127E-09 &	1.163 & 0.604 & 5.232E-07 & 1.369 & 1.107 \\ \hline
8.5 1-,2-,5- T	& 2.900E-05	& 0.670 & 0.240 & 6.000E-06 & 0.760	& 0.300 & 2.260E-05 &	1.713 &	0.474\\ \hline
9.5 1- T  & 1.760E-05 & 0.800 & 0.450 & 7.647E-06 &	1.220 & 0.200 & 1.675E-05 & 1.700 & 0.480  \\ 
 ($\sigma$ = 0.4 MeV) &  & &  &  &	 & & &  &  \\ \hline
9.87 1+, 2+ T & 1.500E-06 & 1.100 & 0.450 & 1.000E-05 & 0.900 & 0.600 & 5.500E-06 & 2.100 & 0.500 \\ \hline
\end{tabular}
\caption{Parameters of our fits of the transverse nuclear excitation form factors (squared) for $\calcium$. The parameterizations are for  $q \equiv |\bf{q_{eff}}|$ in units of fm$^{-1}$.}
\label{excited_states_ca40T}
\end{center}
\vspace{-20 pt}
\end{table*} 
\setlength\extrarowheight{5pt}
\begin{table*}[ht]
\begin{center}
\begin{tabular}{|p{22mm}||p{16mm}|p{11mm}|p{13mm}||p{16mm}|p{11mm}|p{13mm}||p{16mm}|p{11mm}|p{14mm}|}
        \hline
\multicolumn{10}{|c|}{\textbf{$\calcium$ Longitudinal Form Factor Parameters}} \\ \hline
State	&	$N_1$	&	$C_1$	&	$\sigma_1$		&	$N_2$ &	$C_2$	&	$\sigma_2$	&	$N_3$ &	$C_3$	&	$\sigma_3$	 \\  \hline \hline
3.35 0+ L& 5.339E-05 & 0.537 & 0.252 & 1.122E-06 & 1.406 & 0.226 & 5.473E-08 & 2.133 & 0.259 \\ \hline
3.74 3- L& 3.815E-03 & 0.839 & 0.374 & 2.829E-04 & 1.316 & 0.233 & 1.221E-05 & 2.150 & 0.158 \\ \hline
3.90 2+ L&  8.203E-05 & 0.447 & 0.412 & 9.525E-05 & 0.749 & 0.354 & - &	- & - \\ \hline
4.49 5- L& 1.200E-04 & 1.278 & 0.588 & 1.457E-05 & 2.011 & 0.068 & 6.807E-05 &	1.663 &	0.281 \\ \hline
5.25 2+ L& 1.500E-04 & 0.660 & 0.300 &  1.800E-05 & 1.100 & 0.47 & 1.000E-05 & 0.600 & 0.450 \\ \hline
5.61 2+ L& 1.980E-04 & 0.594 & 0.287 & 5.215E-06 & 1.020 & 0.550 & - & - & - \\ \hline
5.90 1- L& 1.000E-04 & 0.670 & 0.350 & 5.800E-05 & 0.850 & 0.350 & 1.500E-06 & 1.000 & 0.500 \\ \hline
6.29 3- L& 8.020E-04 & 0.010 & 0.641 & 1.053E-04 & 0.827 & 0.215 & 1.228E-05	& 1.810 &	0.365 \\ \hline
6.59 3- L& 2.300E-04 & 0.600 & 0.430 & 8.000E-06 & 0.700 & 0.150 & 1.300E-05 & 1.730 & 0.350 \\ \hline
6.90 1- L&  2.251E-03 & 0.655 & 0.296 & 5.755E-04 & 1.028 & 0.257 & 1.164E-05 & 1.915 & 0.278 \\ \hline
7.90 2+,4+ L& 5.251E-04 &	0.658 & 0.234 & 2.367E-05 & 1.115 & 0.172 & 5.691E-05 & 1.115 & 0.578 \\ \hline
8.50  2+,4+ L& 5.200E-04 & 0.003 & 0.569 & 8.771E-05 & 1.221 & 0.280 & 7.546E-06 & 1.828	& 0.224 \\ \hline
\hline 
\end{tabular}
\caption{Parameters for the Longitudinal (L) Form Factors (squared) for the nuclear excited states for $\calcium$. Parameters are for $q \equiv |\bf{q_{eff}}|$ in units of fm$^{-1}$.}
\label{excited_states_ca40L}
\end{center}
\vspace{-20 pt}
\end{table*} 

\begin{figure*}[ht]
\begin{center}
\includegraphics[width=3.4in,height=1.85in]{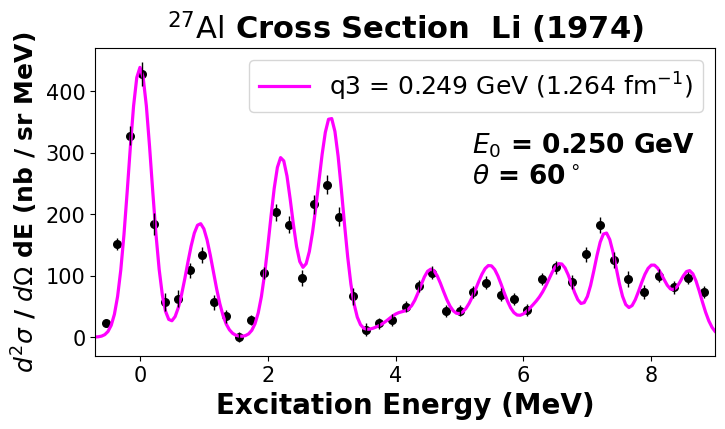}
\includegraphics[width=3.4in,height=1.85in]{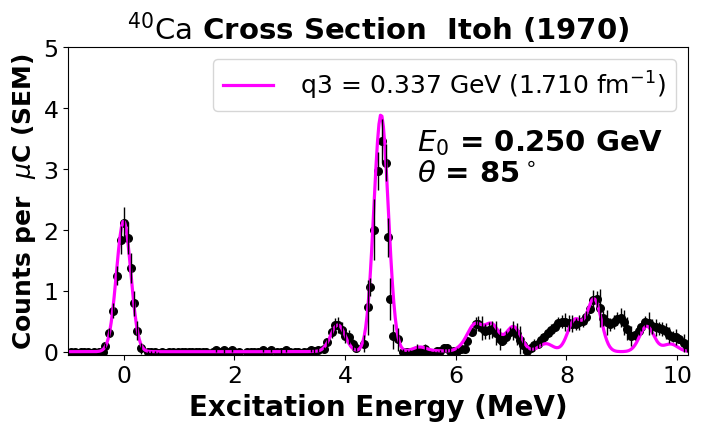}
\caption{{\bf{Left:}} Measurement of cross section \cite{Li:1974vj} (in nb/ sr MeV) for the scattering of 250 MeV electrons from $\aluminum$ at 60 degrees, with a Gaussian smearing resolution of 0.180 MeV. 
{\bf{Right:}} Measurement of cross section \cite{Itoh:1970jb} (in arbitrary units) for the scattering of 250 MeV electrons from  $\calcium$  at 85 degrees, with a Gaussian smearing resolution of 0.130 MeV. The pink solid curve is the predicted cross section using our fits to the form factors (normalized to the elastic peak).}
\label{AL27_comparison}
\end{center}
\end{figure*} 

\setlength\extrarowheight{5pt}
\begin{table*}[b]
\begin{center}
\begin{tabular}
{|p{43mm}||p{15mm}|p{10mm}|p{12mm}||p{15mm}|p{10mm}|p{12mm}||p{15mm}|p{10mm}|p{12mm}|}
        \hline
\multicolumn{10}{|c|}{\textbf{$\aluminum$ Transverse Form Factor Parameters}} \\ \hline
State	&	$N_1$	&	$C_1$	&	$\sigma_1$		&	$N_2$ &	$C_2$	&	$\sigma_2$	&	$N_3$ &	$C_3$	&	$\sigma_3$	 \\  \hline \hline

0.844 1/2+ T	&	7.96E-06	&	0.964	&	0.358	&	2.86E-06	&	1.889	&	0.257	&	2.28E-06	&	2.412	&	0.295	\\ \hline 
1.014 3/2+ T	&	6.40E-06	&	0.82	&	0.319	&	1.58E-06	&	1.527	&	0.743	&	1.29E-06	&	2.044	&	0.234	\\ \hline 
2.210 7/2+ T	&	1.23E-05	&	0.468	&	0.272	&	6.32E-06	&	1.022	&	0.319	&	2.56E-06	&	1.636	&	0.656	\\ \hline 
2.735 5/2+ T	&	1.00E-05	&	1.054	&	0.351	&	8.39E-07	&	1.165	&	0.705	&	1.77E-06	&	2.119	&	0.46	\\ \hline 
2.981 3/2+ T - 3.004 9/2+ T	&	1.02E-05	&	0.443	&	0.266	&	4.96E-06	&	1.011	&	0.381	&	2.48E-06	&	1.886	&	0.619	\\ \hline 
3.680 1/2+ T	&	8.95E-07	&	0.954	&	0.302	&	4.91E-07	&	1.67	&	0.27	&	1.29E-07	&	1.98	&	0.292	\\ \hline 
3.957 3/2+ T	&	4.65E-06	&	0.358	&	0.336	&	2.66E-07	&	1.376	&	0.733	&	8.95E-08	&	2.217	&	0.536	\\ \hline 
4.055 1/2- T	&	1.95E-06	&	1.083	&	0.682	&	2.21E-08	&	2.386	&	1.108	&	4.71E-07	&	1.938	&	0.459	\\ \hline 
4.410 5/2+ T	&	4.00E-07	&	0.5	&	0.6	&	1.00E-05	&	0.8	&	0.5	&	1.00E-06	&	1.7	&	0.4	\\ \hline 
4.510 11/2+ T	&	7.73E-07	&	1.2	&	0.5	&	3.30E-07	&	1.2	&	0.7	&	1.00E-10	&	10	&	1	\\ \hline 
4.580 7/2+ T	&	2.00E-06	&	0.4	&	0.4	&	4.50E-06	&	1.5	&	0.6	&	1.50E-06	&	1.4	&	0.55	\\ \hline 
4.812 5/2+ T	&	7.00E-08	&	0.1	&	0.8	&	3.00E-06	&	0.9	&	0.45	&	1.50E-06	&	1.9	&	0.35	\\ \hline 
5.248 5/2+ T	&	1.46E-05	&	0.637	&	0.363	&	3.88E-06	&	0.808	&	0.432	&	1.41E-06	&	1.633	&	0.559	\\ \hline 
\end{tabular}
\caption{Transverse (T) $\aluminum$  form factors (squared) for nuclear excited states for $\aluminum$.  The states in this table the parameters are for $\bf{q_{eff}}$ in units of fm$^{-1}$.}
\label{excited_states3}
\end{center}
\end{table*}

\subsection {$\aluminum$ and $\calcium$ excitation form factors for  nuclear states with $E_x<$ 10 MeV}

The square of the  longitudinal form factors for the electro-excitation of  $\aluminum$ and  $\calcium$ nuclear excited states are parameterized by the following expressions:

  \begin{equation}
F^2_L(\bf {q_{eff}})=  \bf {q_{eff}}^2  \times \sum_{j=1}^{j=3}  N_j e^{-[(\bf {q_{eff}}-C_j)/\sigma_j]^2}.
\end{equation}

Similarly, for the square of the  transverse form factors for the electro-excitation of  $\aluminum$ and  $\calcium$ nuclear excited states we use the following function: 
\begin{equation}
F^2_T(\bf {q_{eff}})= \bf {q_{eff}} \times \sum_{j=1}^{j=3}  N_j e^{-[(\bf {q_{eff}}-C_j)/\sigma_j]^2},
\end{equation}
where $\bf{q_{eff}}$ is the effective value of the 3-momentum transfer in units of fm$^{-1}$.
\\

The Gaussian parameters for the form factors for the excitation of $\calcium$ states with excitation energies $\textless$ 10 MeV are given in Table \ref{excited_states_ca40T} and \ref{excited_states_ca40L}, while the parameters for the excitation of  $\aluminum$ states with excitation energies $\textless$ 10 MeV are given in Tables \ref{excited_states3} and \ref{excited_states2} in the Appendix. 
\\

Additionally, comparisons of our parameterizations of the $\calcium$ nuclear excitation form factors with experimental data are shown in the left panels of Fig \ref{40Ca_sample_states} and \ref{40Ca_statesL}, and comparisons of our parameterizations of the $\aluminum$ nuclear excitation form factors with experimental data are presented in the  Appendix figures  \ref{AL271}, \ref{AL272}, \ref{AL273}, and \ref{AL274}. Plots of the comparison of form factor measurements to the parameterization for the $\calcium$ 3.74 MeV longitudinal state, and of the ratio to the fit are shown on the right panel of Fig. \ref{40Ca_sample_states}.  These measurements are from (\cite{Heisenberg:1971rw},  \cite{PhysRev.188.1815}) and \cite{Itoh:1970jb}
\\

We note that some of the longitudinal $\calcium$ and $\aluminum$ nuclear excitation form factors are extracted from measurements of total  $|F(q)|^2$\cite{Itoh:1970jb} in combination with  $|F_T (q)|^2$ measurements from \cite{Fagg:1971zz}. Plots of the extracted $|F_L (q)|^2$ for the 5.9, 6.9, 8.5 MeV  $\calcium$ nuclear states, and their corresponding parameterizations, are shown in the bottom panel of Fig \ref{40Ca_statesL}. The $\aluminum$ longitudinal nuclear excitation form factors for the 3.68, 3.957, 4.410, 4.510, 4.580, 4.812, and 5.248  MeV  states are also extracted from measurements of total $|F(q)|^2$ in combination with other $|F_T(q)|^2$ measurements. Detailed plots of the $\aluminum$ extracted form factors are shown in the Appendix. 


\subsection{Comparison to $\calcium$ and $\aluminum$ cross-section data}

We use our $\calcium$ and $\aluminum$ form factor fits to calculate electron scattering cross sections and compare the predictions to experimental electron scattering cross-sections data  for excitation energies less  than 10 MeV.

A Gaussian smearing corresponding to the experimental resolution for each experiment is applied to the cross section prediction for each state. The $\aluminum$ cross-section measurement\cite{Li:1974vj} for  $E_0$ = 0.250 GeV and $\theta$ = 60{\textdegree} is shown on the left panel of Fig \ref{AL27_comparison}. 
The $\calcium$ cross-section counts versus excitation energy measured\cite{Itoh:1970jb} at  $E_0$ = 0.250 GeV and $\theta$ = 85{\textdegree} is shown on the right panel of Fig \ref{AL27_comparison}. Our predictions are shown as the pink solid lines.  

There are no form factor measurements for the $\aluminum$ nuclear states at excitation energies of  7.91, 8.13, 8.38, and 8.5 MeV.  The parameterizations for these states are determined by 
normalizing to the measured cross section for the 7.477 MeV state. The normalization factors of  1.0, 0.5, 0.5, and 1.5 (respectively) are extracted from the cross-section spectra from \cite{Li:1974vj}. We have included the resulting parameterizations in the bottom section of Table \ref{excited_states2} in the Appendix. 
%

\begin{figure*}[ht]
\begin{center}
\includegraphics[width=3.7in,height=4.3in]{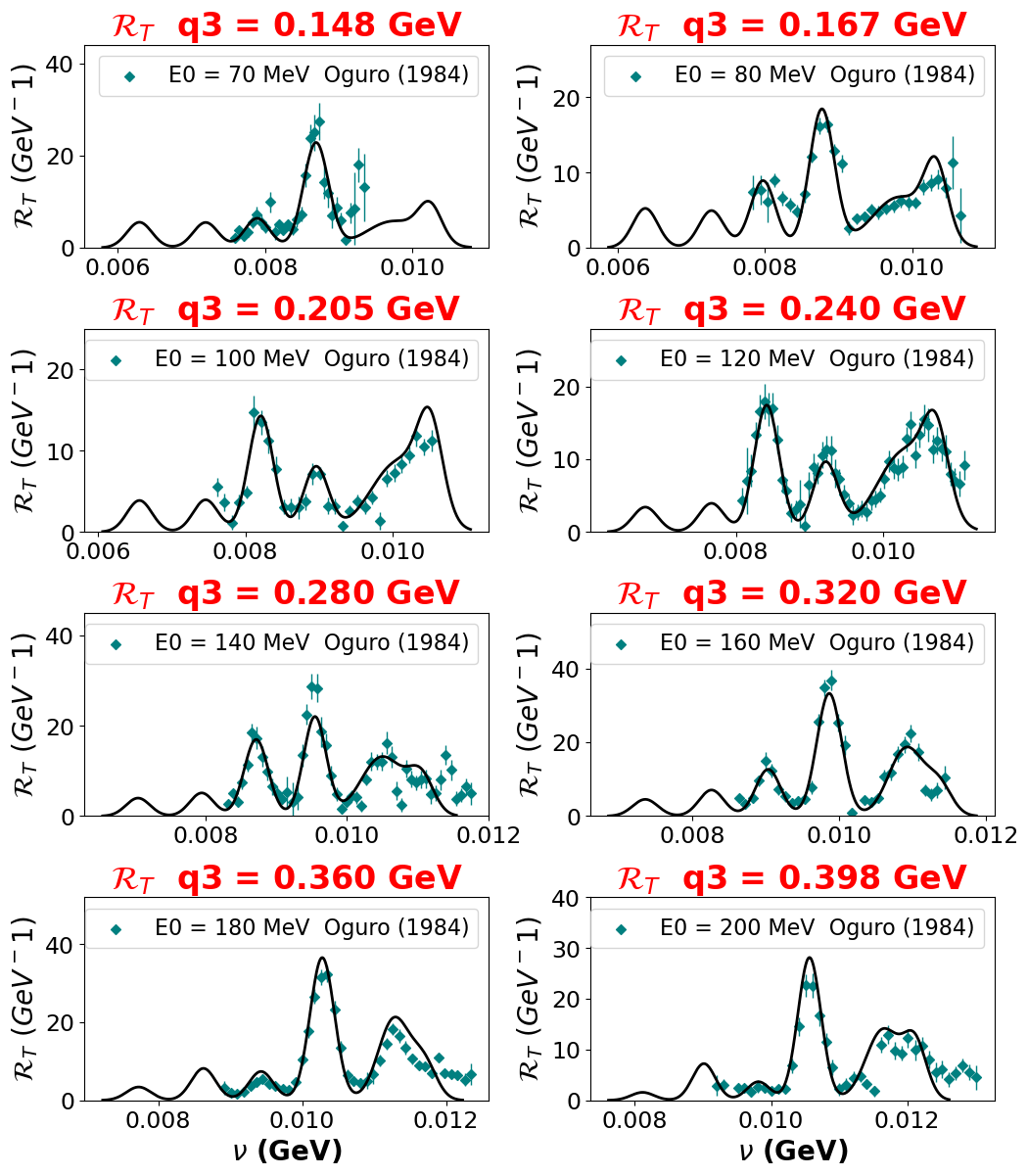}
\includegraphics[width=3.3in,height=4.3in]{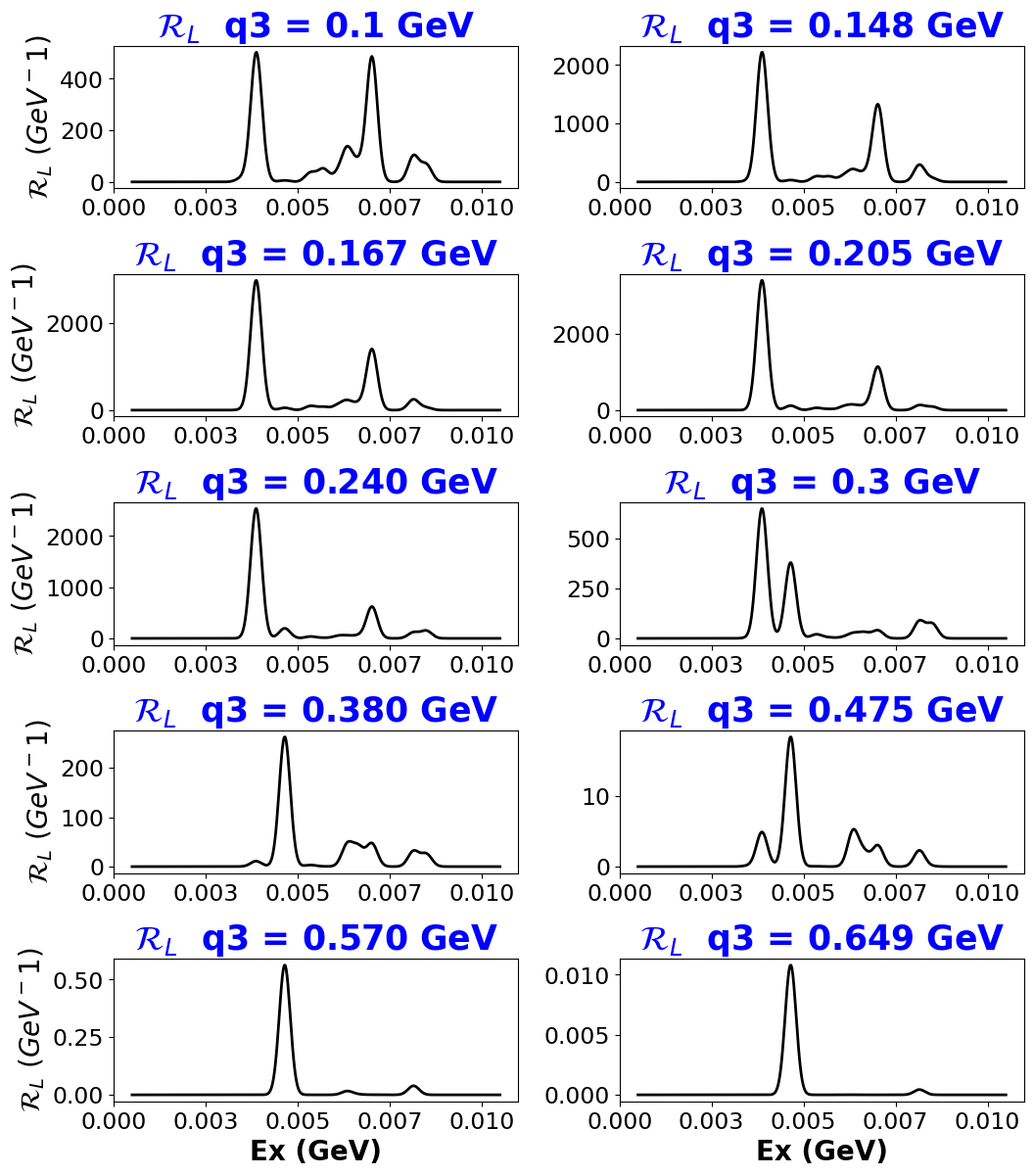}
\caption{{\bf{Left:}} Comparison of the Transverse ${\cal R}_T$ response functions as a function of 3-momentum transfer $\bf{q}$ for $\calcium$ experimental Oguro \cite{Oguro:1984zz} (teal diamond marker) data to the response functions extracted from our parameterizations (solid black line) vs. $\nu$ (GeV) for energies $\textless$ 10 MeV.
{\bf{Right:}} Longitudinal ${\cal R}_L$ response functions as a function of 3-momentum transfer $\bf{q}$ for $\calcium$ extracted from our parameterizations (solid black line) vs. Excitation energy (GeV).}
\label{RTOguro}
\end{center}
\vspace{-20 pt}
\end{figure*} 
%
\begin{figure*}[ht]
\begin{center}
\includegraphics[width=3.3in,height=4.5in]  {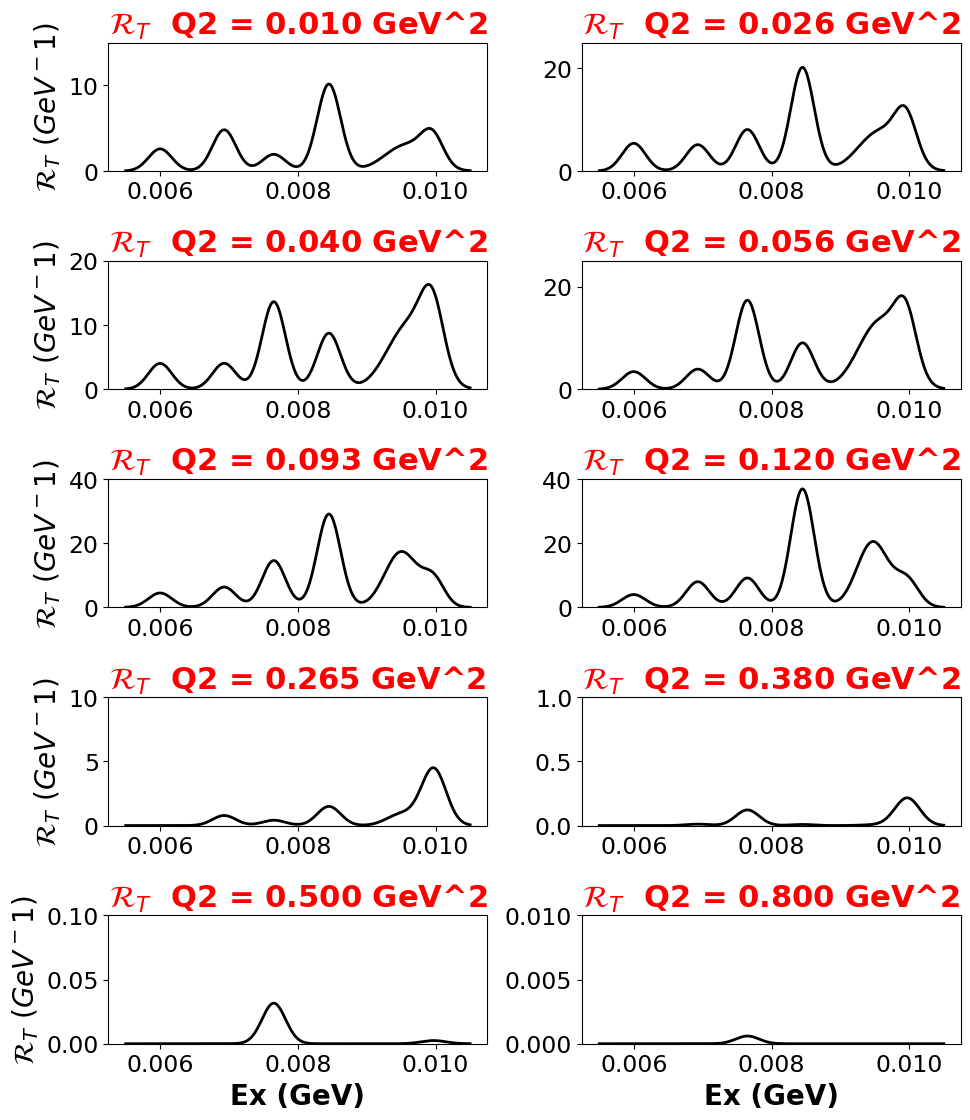}
\includegraphics[width=3.7in,height=4.5in] {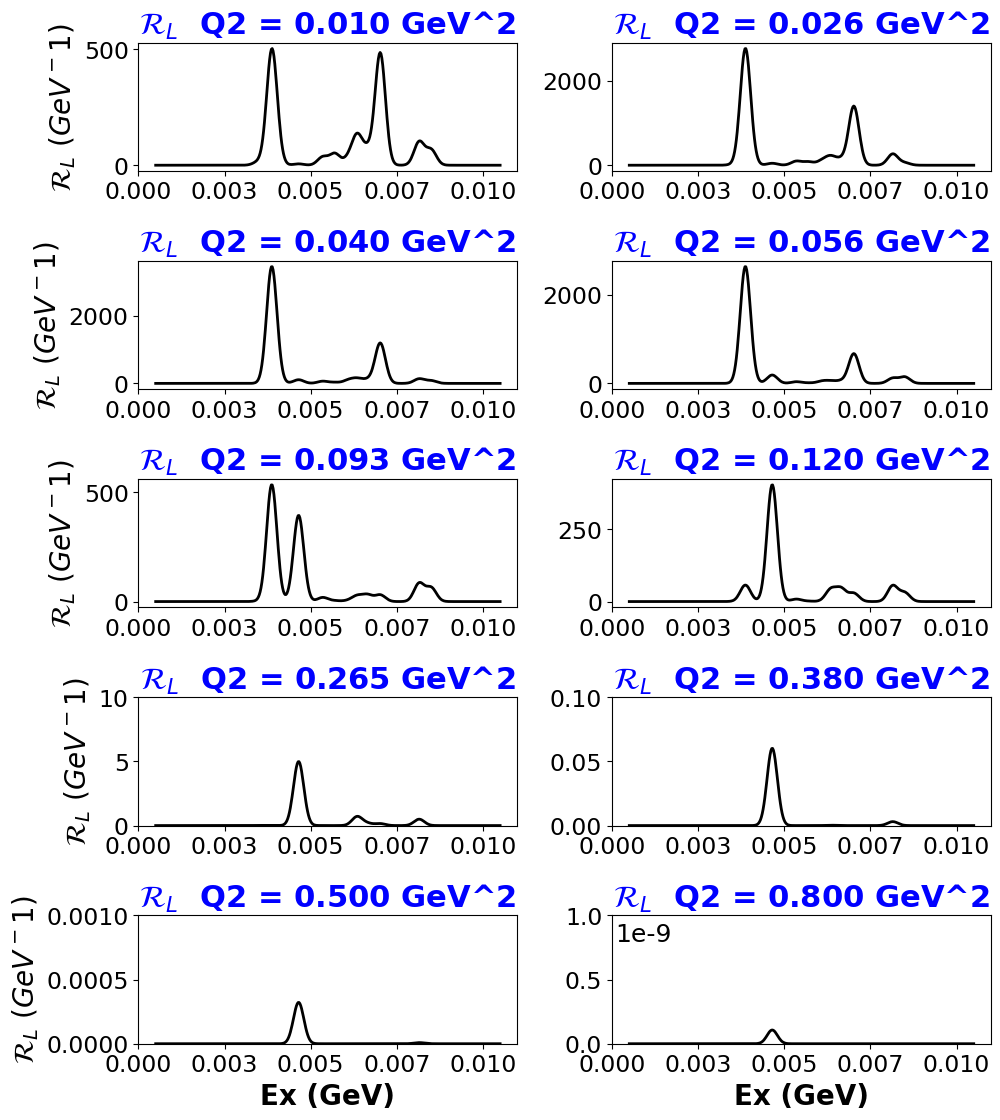}
\caption{{\bf{Left:}} Transverse ${\cal R}_T$ response functions as a function of 4-momentum transfer squared $Q^2$ for $\calcium$ from our parameterizations (solid black line) vs. Excitation Energy (GeV). 
 {\bf{Right}} Longitudinal ${\cal R}_L$ response functions as a function of 4-momentum transfer squared $Q^2$ for $\calcium$ extracted from our parameterizations (solid black line) vs. Excitation energy (GeV).}
\label{RTQ2}
\end{center}
\vspace{-20 pt}
\end{figure*} 
%
\subsection{Predictions of ${\cal R}_L$ and ${\cal R}_T$ data in the nuclear excitation region}

We use our form factor fits to predict ${\cal R}_L$ and ${\cal R}_T$ for $\calcium$ in the nuclear excitation region. A comparison of our predictions for ${\cal R}_T$ to experimental measurements \cite{Oguro:1984zz} are shown in left panels of Fig \ref{RTOguro}. Here the solid black line is our fit based on the nuclear excitation form factor parameterizations. An estimated resolution smearing of 0.2 MeV has been applied to the excitations in the fit to match the data. For the  9.5 MeV T state we applied a different Gaussian smearing resolution of 0.4 MeV to account for the fact that the peak corresponds to  grouping of several adjacent states.

The predicted values of $\calcium$ ${\cal R}_T$ and ${\cal R}_L$ for specific values of $\bf{q}$, and $Q^2$, are shown in Fig \ref{RTOguro} and Fig \ref{RTQ2}, respectively.  These predicted values for excitation energies less than 10 MeV complement other measurements of ${\cal R}_T$ and ${\cal R}_L$ in the Quasielastic region. 

\section{Summary}
We report on empirical parameterizations of longitudinal and transverse nuclear excitation electromagnetic form factors for $\calcium$ and $\aluminum$ for states with excitation energy $\textless$ 10 MeV. We use these fits to predict the  ${\cal R}_T$ and ${\cal R}_L$ response functions as a function of 3-momentum transfer {\bf{q}} and the square of the four momentum transfer $Q^2$ in the nuclear excitation region.
The parameterizations can be used in the testing of theoretical models of electron and neutrino scattering on nuclear targets at low excitation energies. For additonal details see \cite{Amii:2025}.

\section{Acknowledgments}
 Research supported  in part by the Office of Science, Office of Nuclear Physics under contract DE-AC05-06OR23177 (Jefferson Lab) and by  the U.S. Department of Energy under University of Rochester grant number DE-SC0008475.  
\appendix*
\section{Form Factor Parameterizations for $\aluminum$ for $E_x<$ 10 MeV}

\setlength\extrarowheight{5pt}
\begin{table*}[h]
\begin{center}
\begin{tabular}
{|p{43mm}||p{15mm}|p{10mm}|p{12mm}||p{15mm}|p{10mm}|p{12mm}||p{15mm}|p{10mm}|p{12mm}|}
        \hline
\multicolumn{10}{|c|}{\textbf{$\aluminum$ Longitudinal Form Factor Parameters}} \\ \hline
State	&	$N_1$	&	$C_1$	&	$\sigma_1$		&	$N_2$ &	$C_2$	&	$\sigma_2$	&	$N_3$ &	$C_3$	&	$\sigma_3$	 \\  \hline \hline
%
0.844 1/2+ L	&	5.599E-04 & 0.767 & 0.397 & 6.528E-05 & 1.210 & 0.312 & 1.255E-06 & 1.841 & 0.443 \\ \hline
1.014 3/2+ L	&	1.383E-03 & 0.597 & 0.283 & 4.123E-04 & 1.031 & 0.289 &  3.236E-06 & 2.100 & 0.362 \\ \hline
2.210 7/2+ L	&	4.006E-03 & 0.681 &  0.353 & 4.256E-04 & 1.153 & 0.229 & 1.579E-05 & 2.052 & 0.313 \\ \hline
2.735 5/2+ L	&	4.089E-04 & 0.540 & 0.174 & 1.356E-05 & 
1.663 & 0.407 & 3.546E-04 & 0.834 & 0.462 \\ \hline

2.981 3/2+ L - 3.004 9/2+ L & 2.236E-03 & 0.787 & 0.426 & 1.023E-03 & 0.638 &  0.123 & 7.090E-06 & 1.977 & 0.420 \\ \hline
3.680 1/2+ L	&	6.708E-05 & 0.607 & 0.532 & 5.700E-06 & 1.290 & 0.223 & 1.039E-07 & 2.294 & 0.126 \\ \hline
3.957 3/2+ L & 7.287E-05 & 0.325 & 0.335 & 1.230E-06 & 1.221 & 0.663 & 2.134E-08 & 1.808 & 0.225 \\ \hline
4.055 1/2- L & 6.561E-05 & 1.071 & 0.454 & 2.892E-06 & 1.782 & 0.223 & 2.622E-06 & 1.404 & 0.602 \\ \hline
4.410 5/2+  L	& 1.873E-06 & 0.672 & 0.650 & 1.103E-05 & 1.108 &  0.523 & 1.267E-06 & 1.880 & 0.302 \\ \hline
4.510 11/2+ L	& 3.649E-05 & 1.099 & 0.264 & 5.736E-05 & 1.430 & 0.465 & 1.000E-10 & 10.000 & 1.000 \\ \hline
4.580 7/2+ L	&	1.124E-04 & 0.928 & 0.631 & 3.424E-06 & 2.088 & 0.256 &  1.837E-05 & 1.721 & 0.230 \\ \hline
4.812 5/2+  L	&	7.000E-05 & 0.400 & 0.500 & 2.200E-04 & 0.700 & 0.340 & 2.700E-05 & 1.600 & 0.450	\\ \hline
5.156 1/2- L	&	5.082E-04 & 0.035 & 0.524 & 4.019E-05 & 0.869 &  0.691 & 1.292E-05 & 1.190 & 0.223 \\ \hline
5.248  5/2+  L	&	8.000E-07 & 0.550 & 0.086 & 1.550E-04 & 0.650 & 0.315 & 8.859E-06 & 1.676 & 0.382 \\ \hline
5.433 5/2+  L	&	1.754E-03 & 0.235 & 0.446 & 1.481E-04 & 1.038 & 0.337 & 1.096E-05 &  1.465 & 0.597 \\ \hline
5.500 11/2+  L	& 4.999E-05 &  1.253 & 0.488 & 3.072E-06 & 1.894 & 0.240 & 4.530E-07 & 2.414 & 0.258 \\ \hline
5.551 5/2+  L	& 5.841E-07 &  0.213 & 0.414 & 9.812E-05 & 0.654 & 0.354 & 4.162E-06 & 1.341 & 0.310\\ \hline
5.668 9/2+  L	& 8.954E-05 & 0.879 & 0.471 & 1.611E-04 & 0.701 & 0.307 & 2.990E-07 & 2.266 & 0.301 \\ \hline
5.827 3/2- L	&	3.476E-06 & 1.199 & 0.550 & 2.572E-07 & 1.199 & 0.499 & 1.000E-10 & 10.000 & 1.000 \\ \hline
5.960 7/2+  L	&	4.200E-06 & 1.280 & 0.510 & 4.500E-07 & 1.280 & 0.330 & 1.000E-08 & 2.000 & 0.060 \\ \hline
6.159 3/2- L	&	1.136E-04 & 0.878 &  0.481 & 1.148E-05 & 1.348 & 0.078 & 2.217E-07 & 2.320 & 0.137 \\ \hline
6.477 7/2- L	&	1.393E-04 & 0.953 & 0.400 & 1.570E-05 & 1.549 & 0.385 & 5.184E-07 & 2.161 & 0.391 \\ \hline
6.512 11/2+ L - 6.533 9/2+ L 	& 1.741E-05 & 1.005 & 0.371 & 7.497E-06 & 1.352 & 0.616 & 6.500E-07 & 2.350 & 0.350 \\ \hline
6.651 5/2- L	&	2.533E-04 & 0.774 & 0.421 & 5.822E-05 & 1.253 & 0.295 & 2.357E-06 & 1.262 & 1.028 \\ \hline
6.713  9/2+  L	& 3.800E-06 & 1.381 & 0.533 & 1.007E-06 & 1.746 & 0.129 & 4.411E-07 & 2.216 & 0.242 \\ \hline
6.948 11/2+  L	&	2.437E-05 & 0.625 & 0.609 & 1.000E-06 & 1.300 & 0.450 & 5.000E-07 & 2.200 & 0.200	\\ \hline
7.288 9/2- L	&	4.156E-04 & 0.666 & 0.559 &  4.633E-05 & 1.355 & 0.182 & 1.488E-06 & 2.192 & 0.250 \\ \hline
7.477 7/2- L	&	2.519E-04 & 0.975 & 0.320 & 3.783E-05 & 0.892 & 0.596 & 1.717E-06 & 2.223 & 0.267 \\ \hline 
\hline 
7.900 L	&	2.519E-04 & 0.975 & 0.320 & 3.783E-05 & 0.892 & 0.596 & 1.717E-06 & 2.223 & 0.267 \\ \hline 
8.130 L	&	1.259E-04 & 0.975 & 0.320 & 1.892E-05 & 0.892 & 0.596 &  8.585E-07 & 2.223 & 0.267 \\ \hline 
8.376 L	&	1.259E-04 & 0.975 & 0.320 & 1.892E-05 & 0.892 & 0.596 &  8.585E-07 & 2.223 & 0.267  \\ \hline 
8.50 L	&	3.779E-04 & 0.975 & 0.320 & 5.675E-05 & 0.892 & 0.596 & 2.576E-06 & 2.223 & 0.267 \\ \hline 
\end{tabular}
\caption{Longitudinal (L) $\aluminum$  form factors (squared) for nuclear excited states for $\aluminum$.  The  parameters are for $\bf{q_{eff}}$ in units of fm$^{-1}$. Taking into account the lack of data for the nuclear states of 7.91, 8.13, 8.38, and 8.5 MeV, their parametrization have been determined by normalizing with respect to the parametrization of the 7.477 MeV L state.}
\label{excited_states2}
\end{center}
\end{table*}
%
%
\begin{figure*}[h]
\begin{center}
\includegraphics[width=1.70in,height=2.0in]{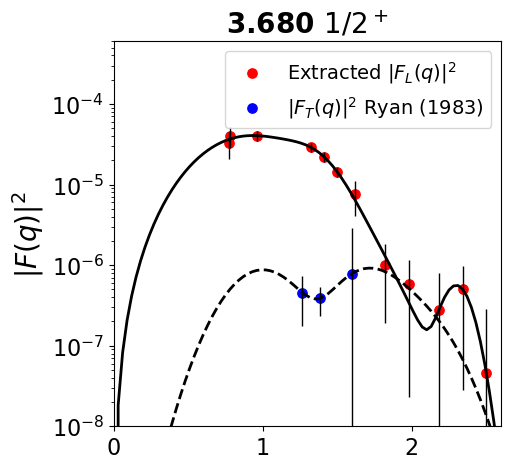}
\includegraphics[width=1.70in,height=2.0in]{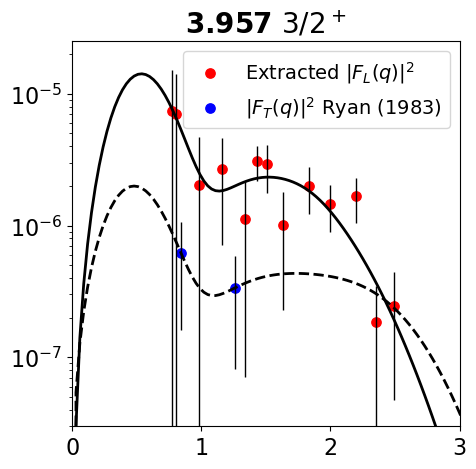}
\includegraphics[width=1.75in,height=2.0in]{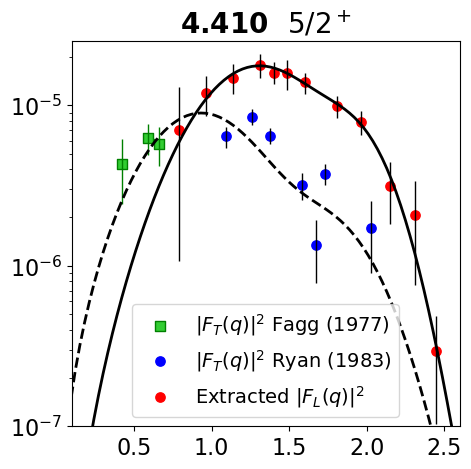}
\includegraphics[width=1.75in,height=2.1in]{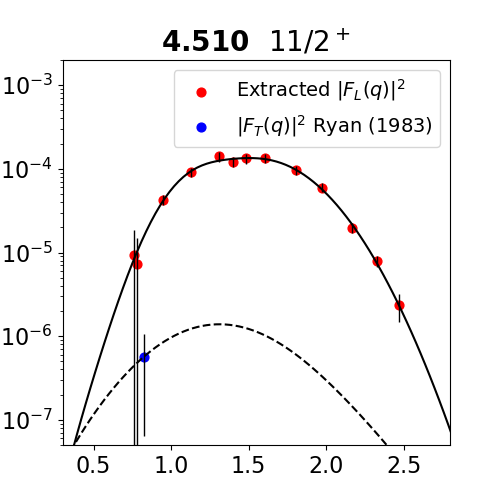}
\includegraphics[width=1.75in,height=2.0in]{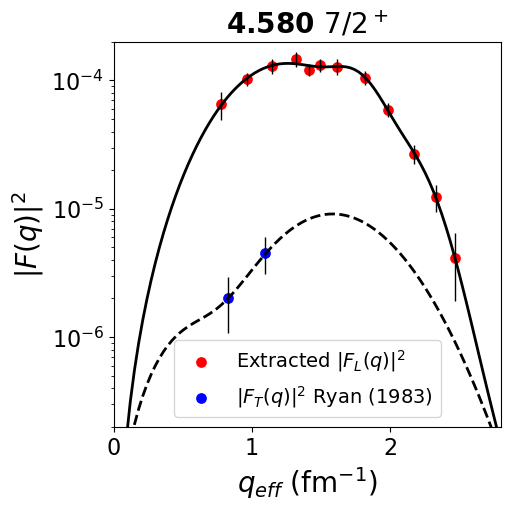}
\includegraphics[width=1.75in,height=2.0in]{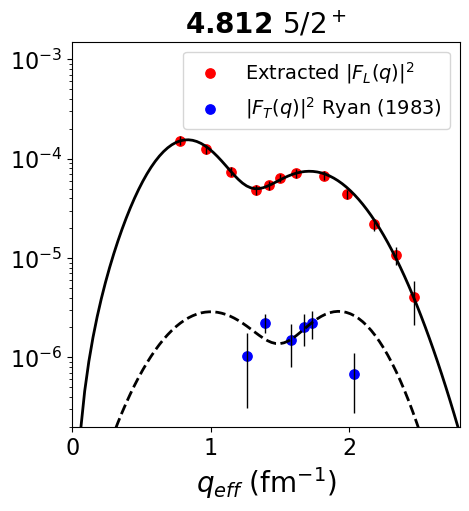}
\includegraphics[width=1.75in,height=2.0in]{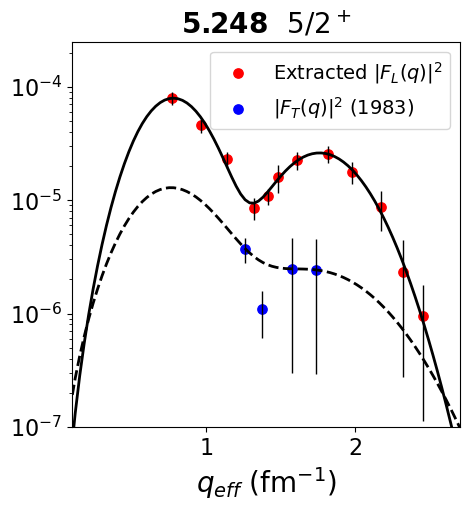}
\caption{Parameterizations of the longitudinal and transverse form factors for $\aluminum$ even parity states (3.680 - 4.812 MeV). For these states, the longitudinal form factors were extracted from experimental total $|F(q)|^2$ and $|F_T (q)|^2$ data in  \cite{Ryan:1983zz}. Blue data points indicate $|F_T (q)|^2$ (parameterizations in dotted-line), while red data points indicate the extracted $|F_L (q)|^2$ (parameterizations in solid-line). Experimental data are taken from \cite{Ryan:1983zz} (blue data points), and \cite{Fagg:1977zz} (green squares).}
\label{AL271}
\end{center}
\end{figure*}

\begin{figure*}[h]
\begin{center}
\includegraphics[width=7.0in,height=1.9in]{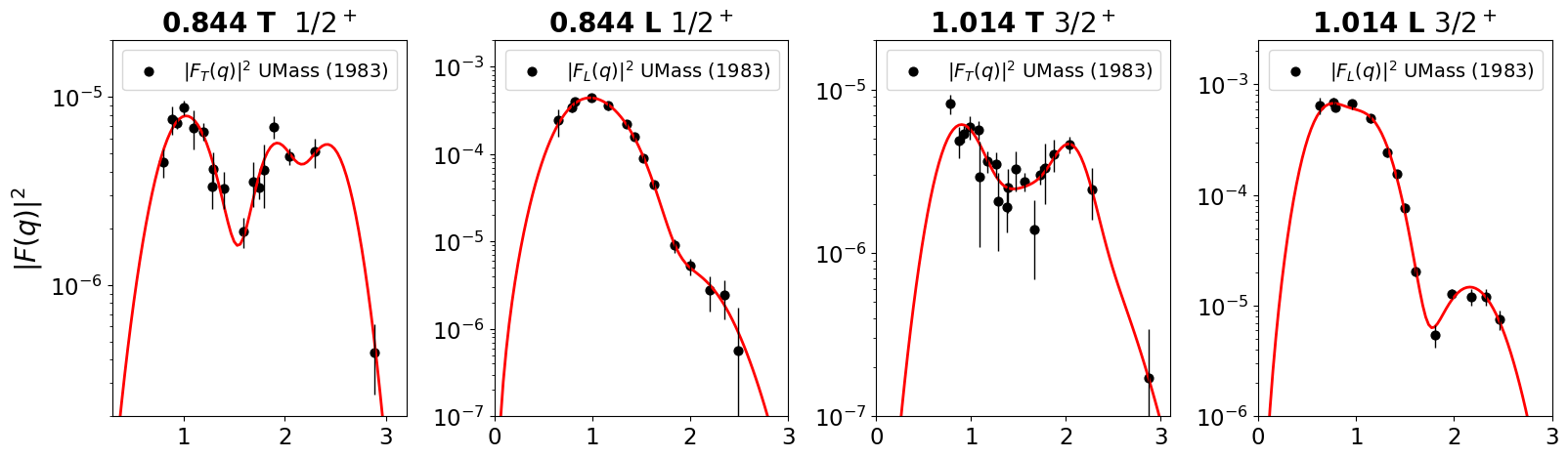}
\includegraphics[width=7.0in,height=1.9in] {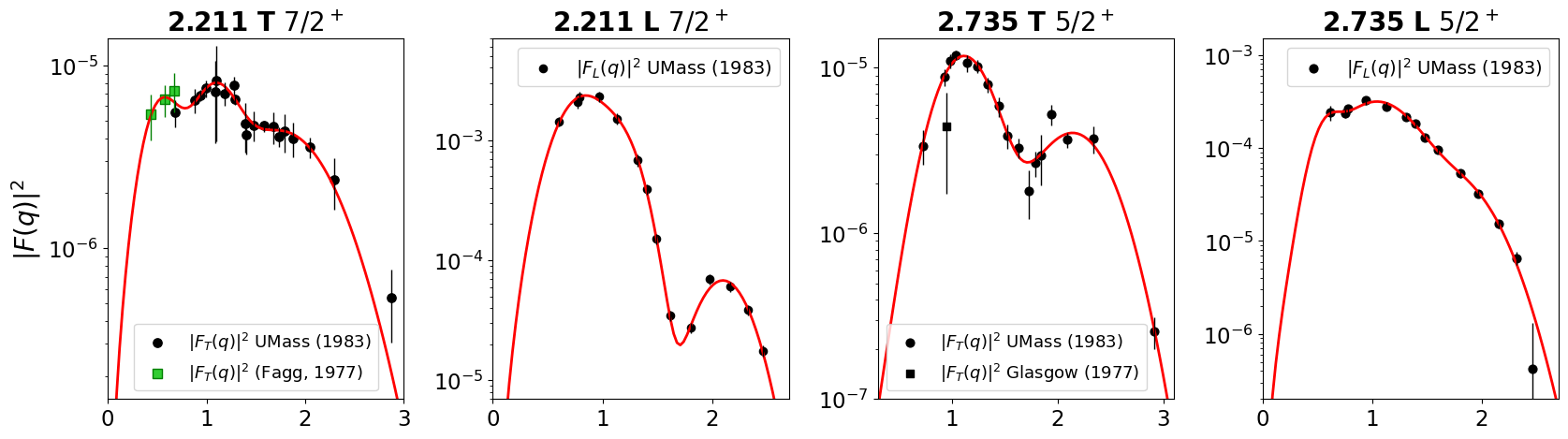}
\includegraphics[width=7.0in,height=2.0in]{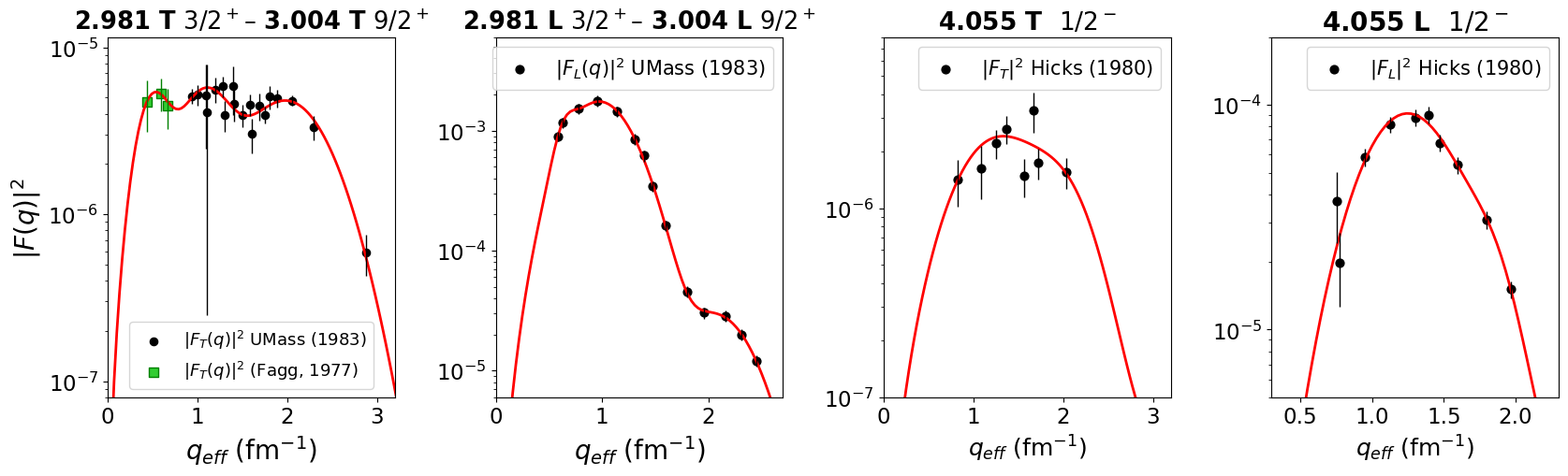}
\caption{Parameterizations of  the longitudinal and transverse form factors for $\rm^{27}Al$ even parity states (0.844 - 3.957 MeV) and the odd parity state (4.055 MeV). Experimental data are taken from \cite{Ryan:1983zz} and \cite{Hicks:1980bv} (black data points), and \cite{Fagg:1977zz} (green squares).}
\label{AL272}
\end{center}
\end{figure*} 

\begin{figure*}[ht]
\begin{center}
\includegraphics[width=7.0in,height=2.0in]{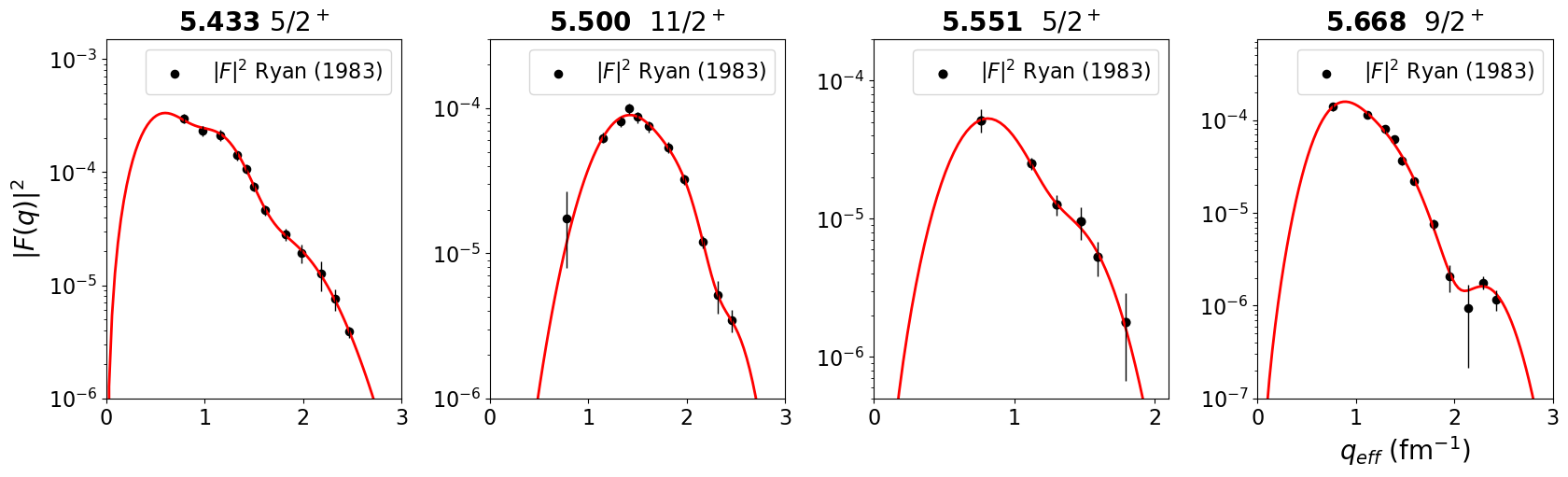}
\includegraphics[width=7.0in,height=2.0in]{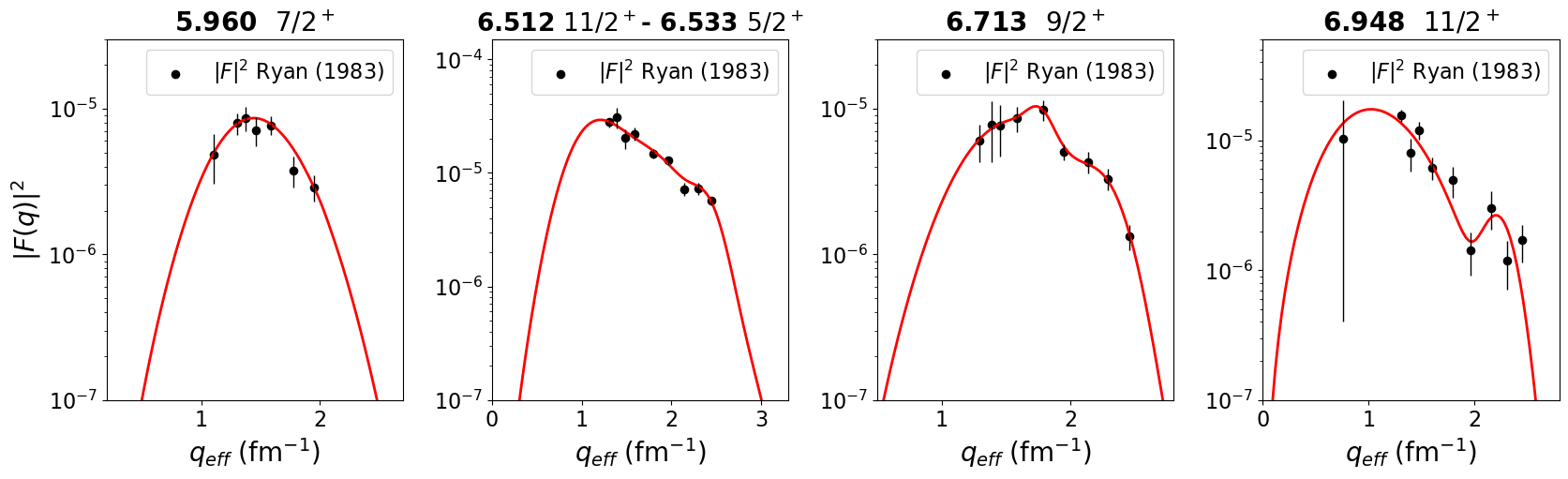}
\caption{Form factors for $\aluminum$ even parity states (5.433 - 6.948 MeV). Experimental data taken from \cite{Ryan:1983zz}}
\label{AL273}
\end{center}
\end{figure*}

\begin{figure*}[ht]
\begin{center}
\includegraphics[width=7.0in,height=1.9in]{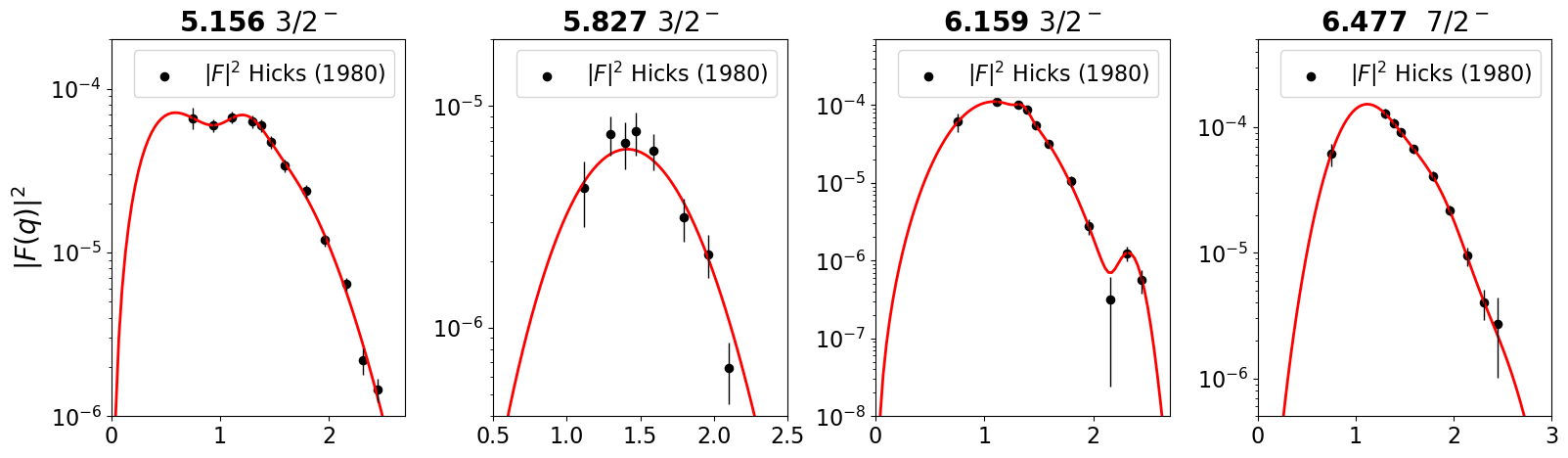}
\includegraphics[width=5.5in,height=2.1in]{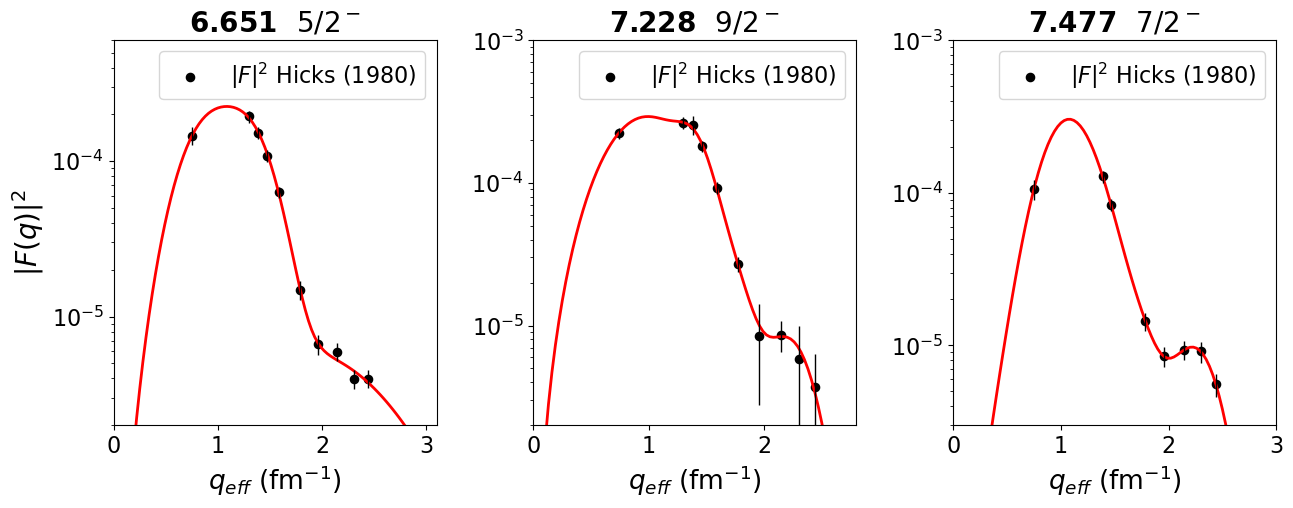}
\caption{Form factors for $\aluminum$ odd parity states (5.156-7.477 MeV). Experimental data taken from \cite{Hicks:1980bv}.}
\label{AL274}
\end{center}
\end{figure*}

\bibliography{AL27_Ca40_states.bib}

\end{document}